\documentclass[printer]{aa}
\usepackage[varg]{txfonts}

\usepackage{graphicx}
\usepackage{natbib}
\usepackage{color}
\usepackage{subfig}
\usepackage[switch]{lineno} 
\usepackage{xspace}
\usepackage{microtype}
\usepackage{amssymb}
\def\link_col{blue}
\usepackage{xspace}
\usepackage{url}
\usepackage{wasysym}
\usepackage{caption}

\def\fermi{{\it Fermi}-LAT \xspace}

\def\grays{$\gamma$-rays\xspace}
\def\gray{$\gamma$-ray\xspace}

\begin{document}

\title{Detection of diffuse \gray emission near the young massive cluster NGC 3603}
\titlerunning{\grays from NGC 3603}
\author{Rui-zhi Yang\inst{1}
\and Felix Aharonian\inst{1, 2}}
\institute{Max-Planck-Institut f{\"u}r Kernphysik, P.O. Box 103980, 69029 Heidelberg, Germany.
\and Dublin Institute for Advanced Studies, 31 Fitzwilliam Place, Dublin 2, Ireland.
}%
\date{Received:  / Accepted: } 

\abstract
{We report the Fermi Large Area Telescope's detection of  extended \gray emission towards the direction of the young massive star cluster NGC 3603. The emission shows a hard spectrum with a photon index of 2.3 from 1~GeV to 250~GeV. The large size and high luminosity of this structure make it unlikely a pulsar wind nebular. On the other hand the spatial correlation with the ionised gas indicate a hadronic origin. The total cosmic ray (CR) protons energy are estimated to be of the order $10^{50} ~\rm erg$ assuming the \grays are produced in the interaction of CRs with ambient gas . The environment and spectral features show significant similarity with the Cygnus cocoon. It reveals that the young star clusters may be a \gray source population and  they can potentially accelerate a  significant fraction of  the Galactic cosmic rays.}
\maketitle
\keywords{Gamma rays: ISM -- (ISM:) cosmic rays }

\section{Introduction}
The current paradigm of cosmic rays (CRs) postulates that the acceleration sites of CRs are supernova remnant (SNRs).  However,   the recent measurements of $^{60} Fe $  abundance in CRs \citep{binns16} indicate that a substantial fraction of CRs could be accelerated in  young OB star clusters and related super bubbles. Furthermore, the measurements of the Galactic diffuse \gray emission   shows that the CRs have a similar radial distribution as OB stars rather than SNRs \citep{fermi_diffuse,yang16}.  On the other hand, super bubbles do have  sufficient kinetic energy, supplied by supernova explosions therein or collective stelar winds, to provide the flux of locally measured CRs\citep{parizot04}.  Meanwhile, thees objects should be visible in \grays due to the fresh accelerated CRs interacting with ambient gas. So far the only detection of such sources is Cygnus cocoon. One should note that Cygnus OB system is by far not the most powerful young star cluster in our Galaxy. 

To search other analogues we have chosen NGC 3603, which is densest young star cluster in our Galaxy. NGC 3603 locates about 7 kpc away from the solar system and it is the Galactic clone of the core of 30 Doradus in Large Megelanic cloud. Also, it  is one of the most massive \ion{H}{ii} regions in our Galaxy \citep{moffat94}. The age of NGC 3603 is about 2 Myr \citep{kudryavtseva12}. It contains more than 50 OB stars and Wolf-Rayet stars in a very compact region. These stellar winds from massive young stars provide sufficient kinetic energy to accelerated CRs.  The weak non-thermal x-ray emission has been detected in this region as well \citep{moffat02}. 

We have structured the paper as follows: in Section~2, we present the results of our analysis of the \fermi observations and discuss the observational uncertainties of the used procedure; In Section 3 we discuss the origin of the gamma-ray emission. We estimate the mass of gas in the based on the dust opacity maps, and derive the CR spectra and fluxes assuming that gamma-rays are produced in interactions of CR protons and nuclei  with the ambient gas. In Section  4 we discussed other possible origin of the \gray emission; in Section~5 we present the main conclusion of this work. %

\section{Fermi LAT data analysis}
\subsection{Spatial analysis}
We selected observations for which the \fermi detector was pointed towards NGC 3603 (MET 239557417 -- MET 455067824), for a period of approximately 7 years.
For this analysis, we have used the standard LAT analysis software package \emph{v10r0p5}\footnote{\url{http://fermi.gsfc.nasa.gov/ssc}}. 
Given the crowded nature of the region and to avoid systematic errors due to the poor angular resolution at low energies, as well as the domination of pulsar emission below several GeV in the Galactic plane,  we selected only events with energies exceeding 10~GeV for the spatial analysis. 
The region-of-interest (ROI) was selected to be a $15^ \circ \times 15^ \circ$ square centred on the position of NGC 3603, i.e., ra=$168^{\circ}.848$, dec=$-61^{\circ}.251$
In order to reduce the effect of the Earth albedo background, we excluded from the analysis the time intervals when the Earth was in the field-of-view (specifically when the centre of the field-of-view was more than $52^ \circ$ from zenith), as well as the time intervals when parts of the ROI had been observed at zenith angles $> 100^ \circ$. 
The spectral analysis was performed based on the P8\_R2\_v6 version of the  post-launch instrument response functions (IRFs). Both the front and back converted photons were selected.

\begin{table*}[htbp]
\caption{Fitting results for different models} \label{tab:loglike} \centering
\begin{tabular}{llll}
\hline
Model &\vline ~-log(likelihood)&\vline ~TS for the extended structure\\
\hline
diffuse model + 3FGL catalog   &\vline ~30637 &\vline\\

\hline
diffuse model + 3FGL catalog + Gaussian disk   &\vline  ~30580 &\vline~114\\
\hline
diffuse model + 3FGL catalog + five point sources   &\vline  ~30613 &\vline\\
\hline
\end{tabular}
\end{table*}

\begin{figure*}
\centering
\includegraphics[width=0.3\linewidth]{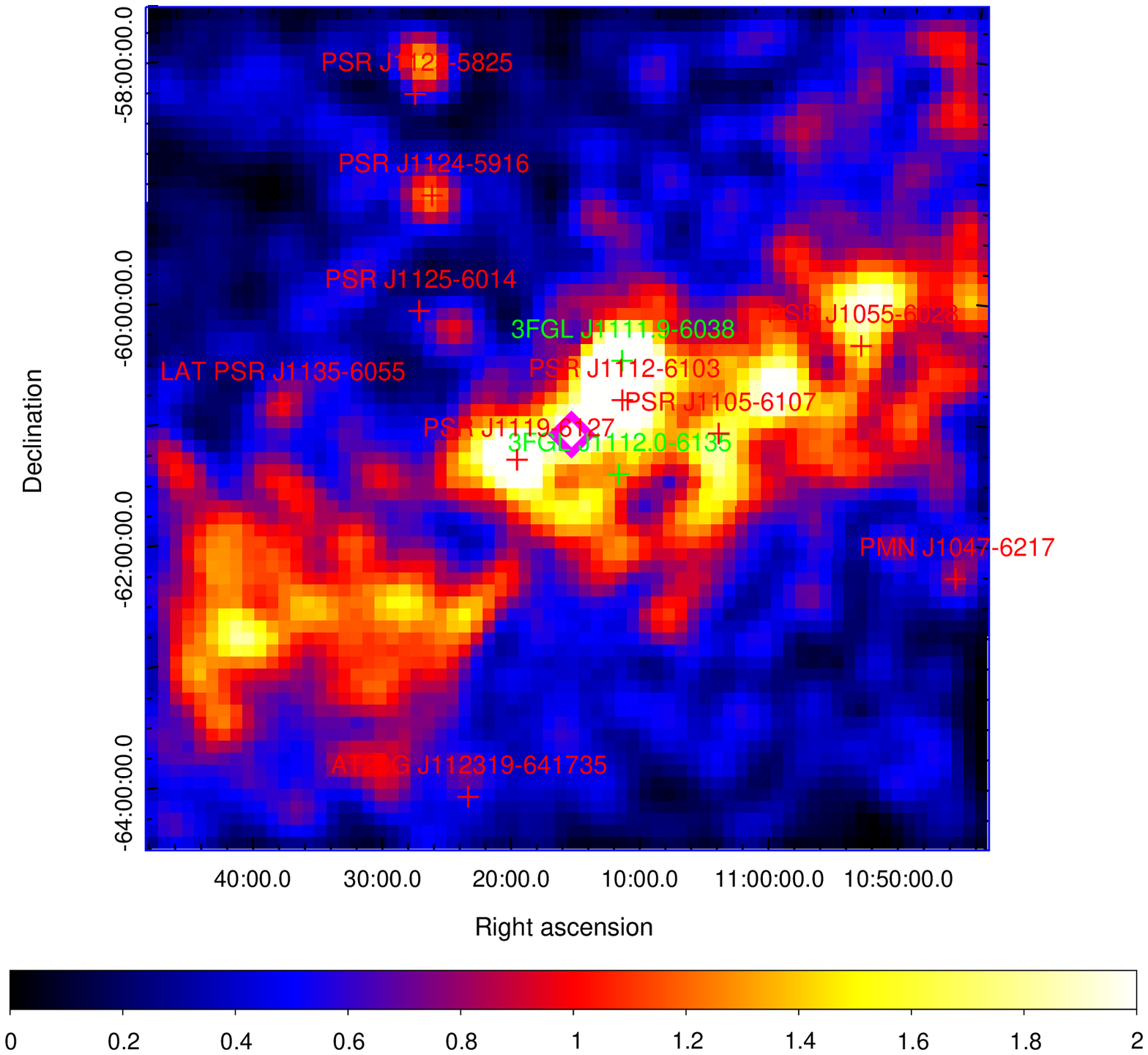}\includegraphics[width=0.3\linewidth]{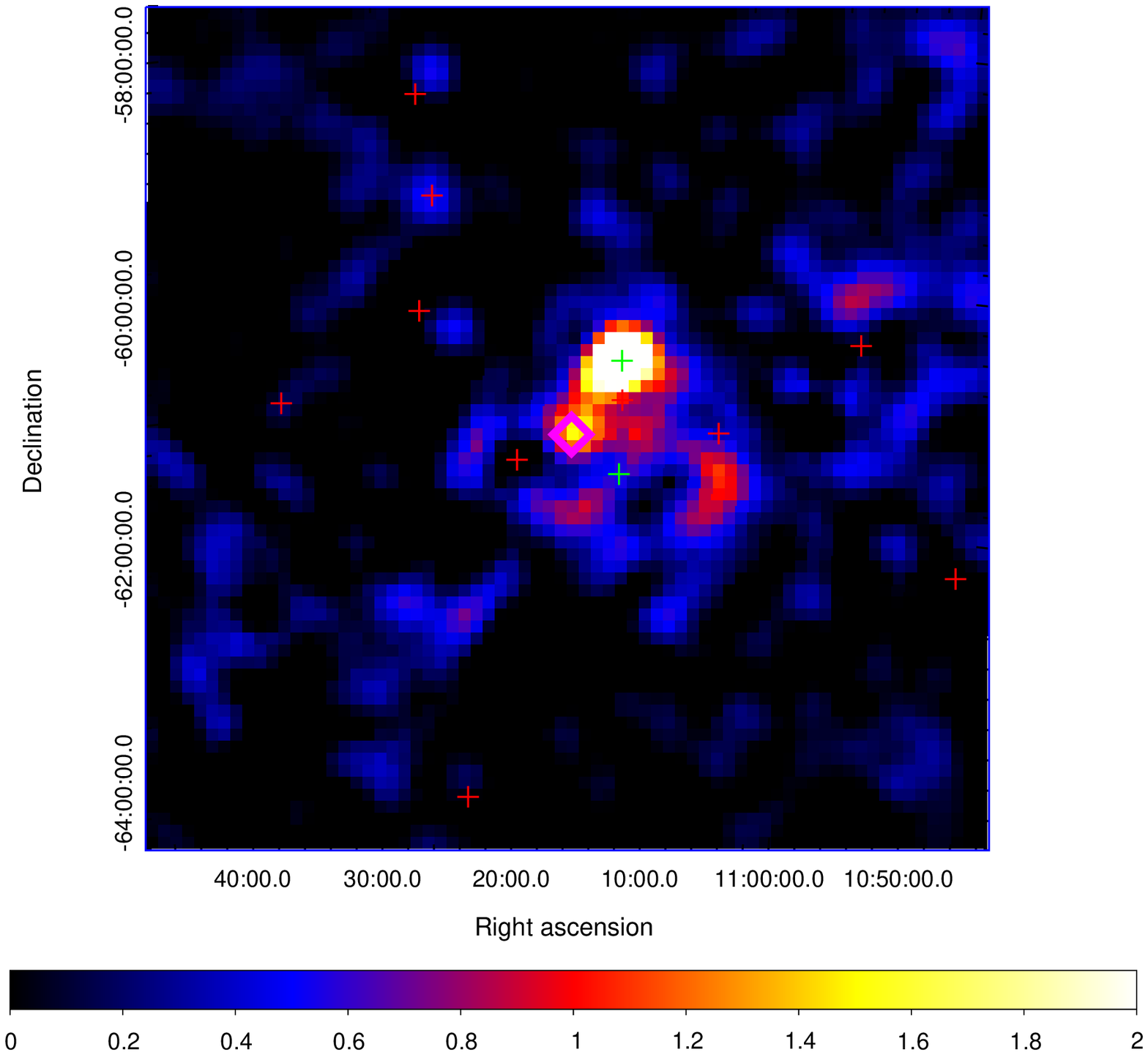}\includegraphics[width=0.3\linewidth]{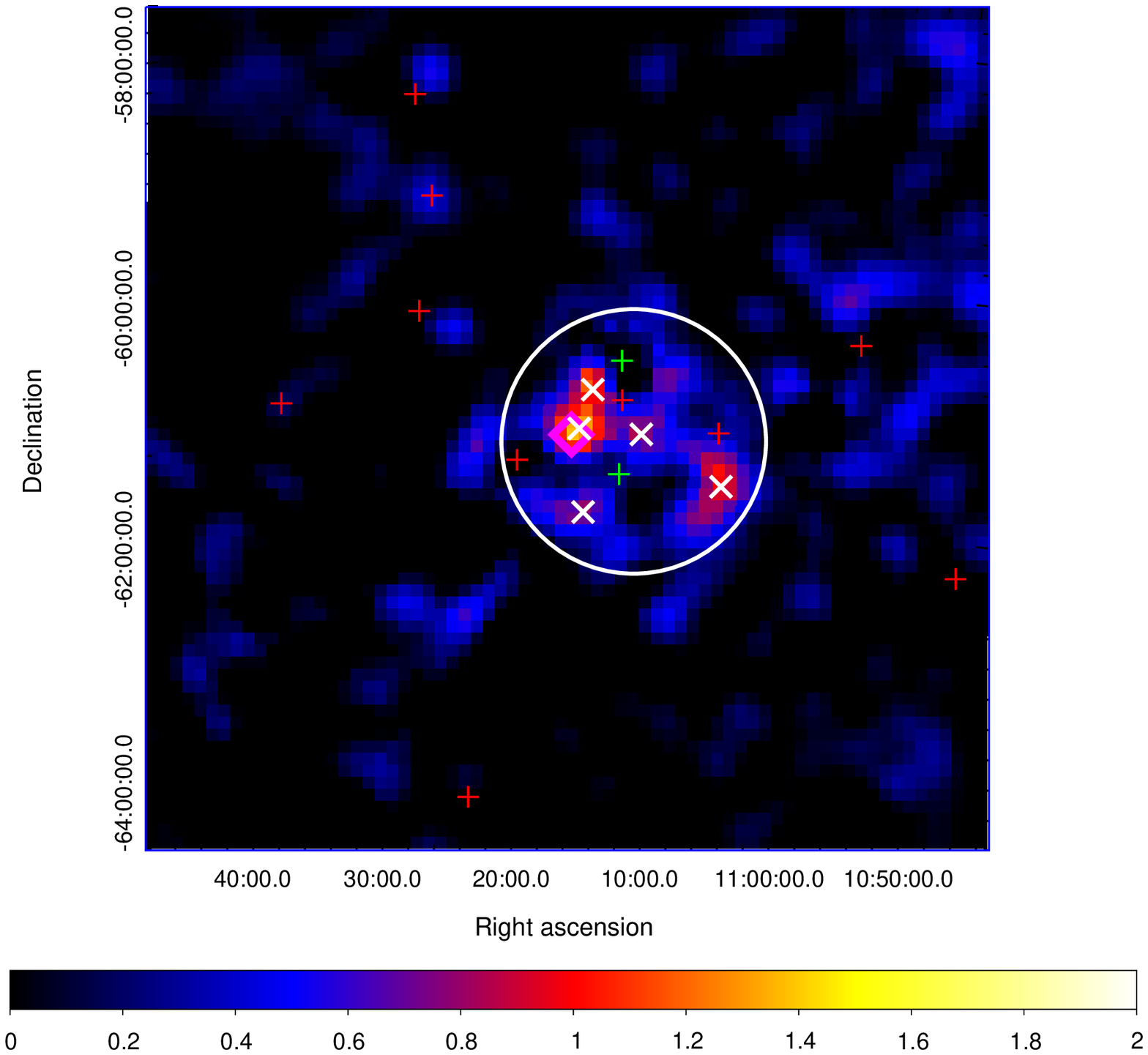}
\caption{{\it left} \grays counts map above 10~GeV in the inner $5^{\circ}$ around NGC 3603. The identified 3FGL catalog sources are labeled as red crosses. The two unassociated catalog sources are labeled as green cross. The position of NGC 3603 is marked as a magenta diamond.  {\it middle} : The residual map after subtracting all the identified catalog source and the diffuse background. {\it right} :  The residual map after subtracting all the identified catalog source and the diffuse background, as well as the unassociated catalog source 3FGL 1111.9-6038. Also shown is the best fit gaussian disk (white circle, the radius is corresponding to the $1~\sigma$ of the Gaussian) and the position of the five point sources (white "x") used to test the hypothesis that the extended emission comes from several independent point sources.   }
\label{fig:skymap}
\end{figure*}

The \gray counts map above 10 GeV of the inner $5^{\circ}$ is shown in the top left panel of  figure.\ref{fig:skymap}. The identified Fermi LAT point sources listed in the 3rd Fermi source catalog (3FGL)\citep{3fgl} are also shown as red crosses.  In this region pulsars dominate the point source population. There are also two unassociated Fermi point sources near NGC 3603, which are marked as green crosses in the figure.  To derive the  emission toward NGC 3603 we perform a likelihood analysis by using the tool {\it gtlike}.  In the likelihood fitting we first include all the sources in the 3FGL catalog and the fermi diffuse background model gll\_iem\_v06.fits for the Galactic \gray emission, as well as the isotropic background model iso\_P8R2\_SOURCE\_V6\_v06.txt\footnote{ available at \url{http://fermi.gsfc.nasa.gov/ssc/data/access/lat/BackgroundModels.html}}.  The normalization and spectral index of the sources are left free in the analysis. 

After the likelihood fitting we subtracted the best fit diffuse model and all the identified sources in the ROI, the resulted residual maps are shown in top right panel of figure. \ref{fig:skymap}. We found  strong residuals towards the direction of NGC 3603 (marked as purple diamond) and the unassosicated Fermi source 3FGL 1111.9-6038. \citet{slane12} reveals that this source 3FGL 1111.9-6038 should be related to the supernova remnant MSH 11-62.  The other unassosicated Fermi source 3FGL 1112.0-6135, is not significant above 10~GeV and thus has little influence on the results. We then further subtracted the source 3FGL 1111.9-6038 and the residuals are shown in the bottom panel of Figure. \ref{fig:skymap}. A diffuse emission peaks at the position of NGC 3603. 

To study the morphology of the diffuse emission, we added a gaussian disk on top of the model used in the likelihood analysis. We then vary the position and size of the disk to find the best fit parameters. The best-fit result is a gaussian disk centered at (ra=$167^{\circ}.78$, dec=$-61^{\circ}.28$) with $\sigma=1.1^{\circ}$, with a TS value of 114, corresponding to a significance of more than 10 $\sigma$.   We also test whether this extended emission is composed by several independent point sources.  To do this we added 5 point source at the peaks in the residual maps. These point sources are not significant except for the one which coincide in position with NGC 3603. And the -log(likelihood) function value is larger than the gaussian disk case, even with more free parameters. Thus the point source scenario can be significantly excluded.  We also note that the morphology of the residual reveal the hint for deviation from a simple gaussian disk, but the limited statistics prevent us to pursue this issue deeper. Thus in the following analysis we use the best-fit gaussian disk as the spatial template.   We list the model and the -log(likelihood) value in Tab.\ref{tab:loglike}. 

\subsection{Spectral analysis}
\begin{figure*}
\centering
\includegraphics[width=0.8\linewidth]{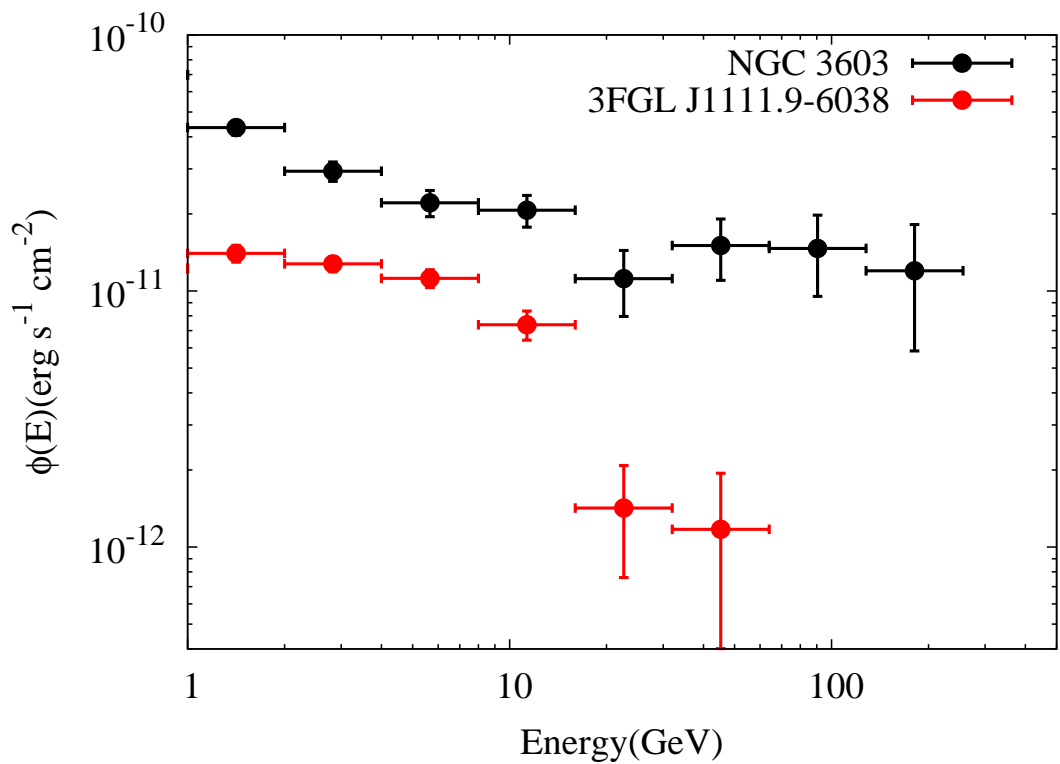}
\caption{The SEDs of the extended emission towards NGC 3603 and 3FGL 1111.9-6038.
}
\label{fig:SED}
\end{figure*}

To obtain the spectral energy distribution (SED) of extended emission towards NGC 3603, we divided the energy range $1000~ {\rm MeV} - 250~{\rm GeV}$ into 8 logarithmically spaced bands and applied \emph{gtlike} to each of these bands. 
The results of this analysis are shown in Figure~\ref{fig:SED}. All data points have a test statistic (TS) values larger than 4, which corresponds to a significance of greater than $2\sigma$. The spectral points can be well fitted with a power law with a photon index of 2.3. The power law spectrum extends to 250 GeV without any sign of cutoff. The total \gray luminosity above 1~GeV is $10^{36}~\rm erg$ given the distance of 7~kpc. 
In addition, we also plot the SED for the source  3FGL 1111.9-6038, which reveals a break above about 20 GeV. 

\section{CR accelerated in young star cluster }

The young star cluster NGC 3603 contains more than 50 OB stars and Wolf-Rayet stars \citep{moffat94}, the collective wind or historical supernova explosion inside the cluster make it a potential accelerator of CRs\citep{parizot04}. The hard \gray emission has already been detected in the young star association Cygnus OB2 \citep{fermi_cygnus}. NGC 3603 has more OB stars and a much denser star distribution compared to Cygnus OB2, which provide enough kinetic energy in the stellar wind and make it a even more powerful acceleration site. Thus a natural explanation of the extended \grays are the fresh accelerated CRs illuminating the ambient gas. 

The gas content can be investigated by various tracers. The traditional tracers of the hydrogen in the atomic and   
molecular forms are the  21 cm \ion{H}{i}  and 2.6 mm CO lines, respectively.  In this paper we use the data from CO galactic survey of  \citet{dame01} with the 
CfA 1.2m millimetre-wave Telescope,  and the Leiden/Argentine/Bonn (LAB) Survey on \ion{H}{i} gas. 
For the CO  data, we use the standard assumption of a linear relationship between the velocity-integrated 
CO intensity, $W_{\rm CO}$, and the column density of molecular hydrogen, N(H$_{2}$). 
The conversion factor $X_{\rm CO}$ is chosen to be $2.0\times10^{20} ~\rm cm^{-2} (K~km~s^{-1})^{-1}$ as suggested by \citet{dame01, bolatto13}. 
For the \ion{H}{i} data  we use the equation 
\begin{equation}
N_{\ion{H}{i}}(v,T_s)=-log \left(1-\frac{T_B}{T_s-T_{bg}}\right)T_sC_i\Delta v \ ,
\end{equation}
where $T_{bg}\approx2.66$~K is the brightness temperature of the cosmic microwave background radiation at 21cm, and
$C_i = 1.83 \times 10^{18} \rm  cm^{2}$.  In  the case  when $T_B > T_s-5~\rm K$, 
we truncate $T_B$ to $T_s-5~\rm K$; 
$T_s$ is chosen to be 150 K. The systematic uncertainties due to the different spin temperatures are discussed in \citet{fermi_diffuse_old} and \citet{fermi_diffuse} and the effect  is  quite small in most regions of the sky.   \citet{hummel15} has revealed that for NGC 3603 most of the CO emission fall into the velocity of $0 - 20~\rm km/s $. We also use this range to integrate the line emission both of CO and 21 cm in this velocity range.

For  different  reasons,  the neural gas  cannot  be always  traced by  CO and \ion{H}{i} observations \citep{grenier05}. 
In such  cases (e.g. in optically thick clouds),  the  infrared emission from cold interstellar 
dust provides an alternative and  independent  measurements  of the gas column density.  
To find  it, we need a relation between the dust opacity and the column density.  
According to Eq.~(4) of \citet{planck}, 
\begin{equation}\label{eq:dust}
\tau_M(\lambda) = \left(\frac{\tau_D(\lambda)}{N_H}\right)^{dust}[N_{H{\rm I}}+2X_{CO}W_{CO}],
 \end{equation}
where $\tau_M$ is the dust opacity as a function of the wavelength  $\lambda$,  $(\tau_D/N_H)^{dust}$ is the reference dust emissivity measured in low-$N_H$ regions, $W_{CO}$ is the integrated brightness temperature of the CO emission, and $X_{CO}=N_{H_{2}}/W_{CO}$ is the  $H_2/CO$ conversion factor.
The substitution of  the latter into Eq.~(\ref{eq:dust})  gives 
\begin{equation}
N_H = N_{H{\rm I}} +2 N_{H_2} =  \tau_m(\lambda)\left[\left(\frac{\tau_D(\lambda)}{N_H}\right)^{dust}\right]^{-1}. 
\end{equation}
Here  for the dust emissivity at $353~\rm GHz$,   we use  $(\tau_D/N_H)^{dust}_{353{\rm~GHz}}=1.18\pm0.17\times10^{-26}$~cm$^2$  taken from Table~3 of  \citet{planck}. 
Generally,  the dust opacity is considered  as a robust and reliable estimate of the gas column density. 
On the other hand, the dust opacity maps do not contain any  information on the distance.  Thus the column derived from dust opacity is the integration over the whole line of sight and should be regarded as an upper limit.

NGC 3603 is the most massive \ion{H}{ii} region in the Galaxy.  To determine the \ion{H}{ii} column we used the planck free-free map\citep{planck15-10}. We first converted the emission measure (EM) in the planck map into free-free intensity by using the conversion factor in Table 1 of \citet{finkbeiner03}. Then we used Eq.(5) of \citet{sodroski97} to calculate the \ion{H}{ii} column from free-free intensity. We noted that the derived  \ion{H}{ii}  column is inverse proportional to electron density $n_e$, which are chosen to be $2 ~\rm cm^{-3}$ and $10 ~\rm cm^{-3}$  as an upper and lower limit here. 

\begin{figure*}
\centering
\includegraphics[width=0.3\linewidth]{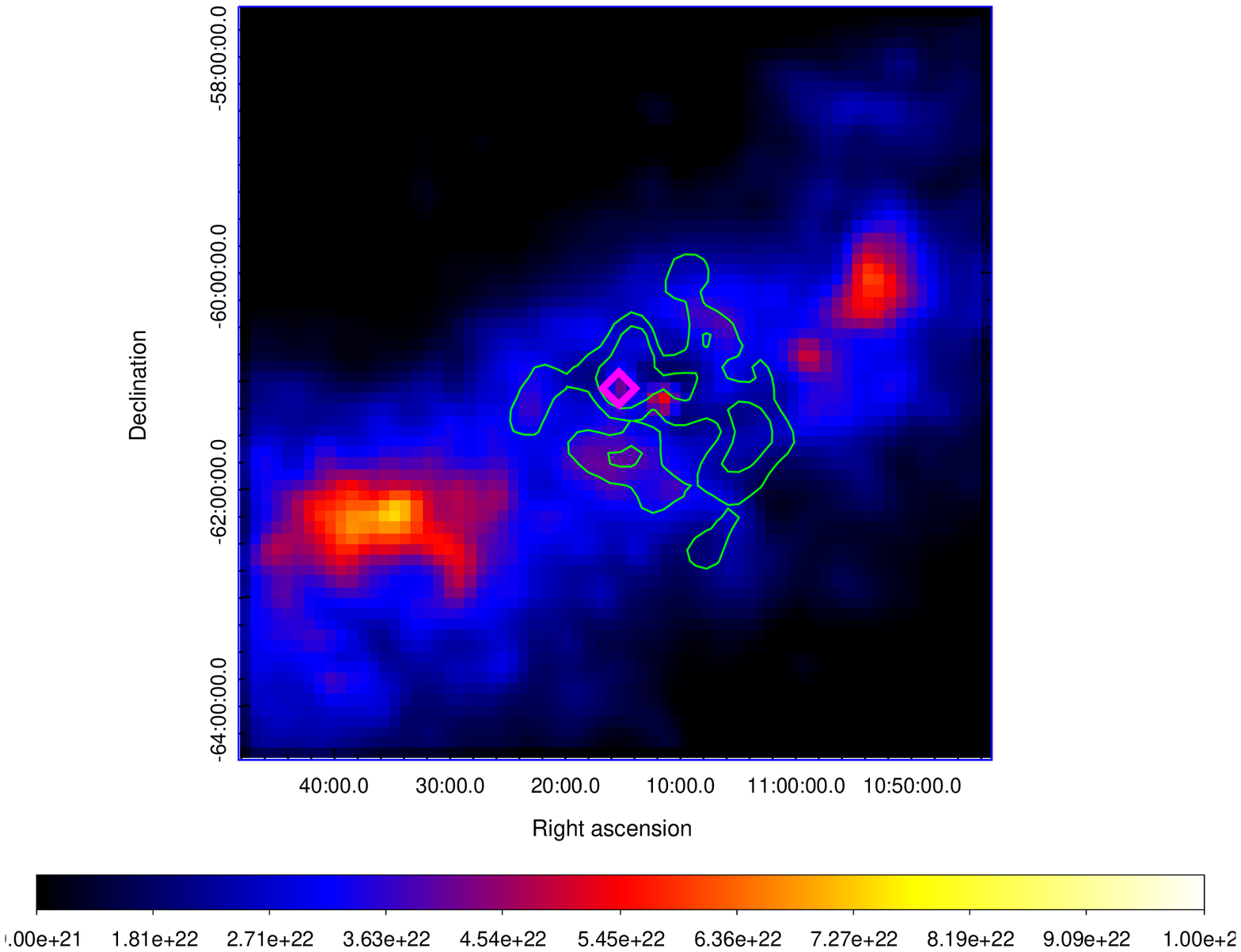}\includegraphics[width=0.3\linewidth]{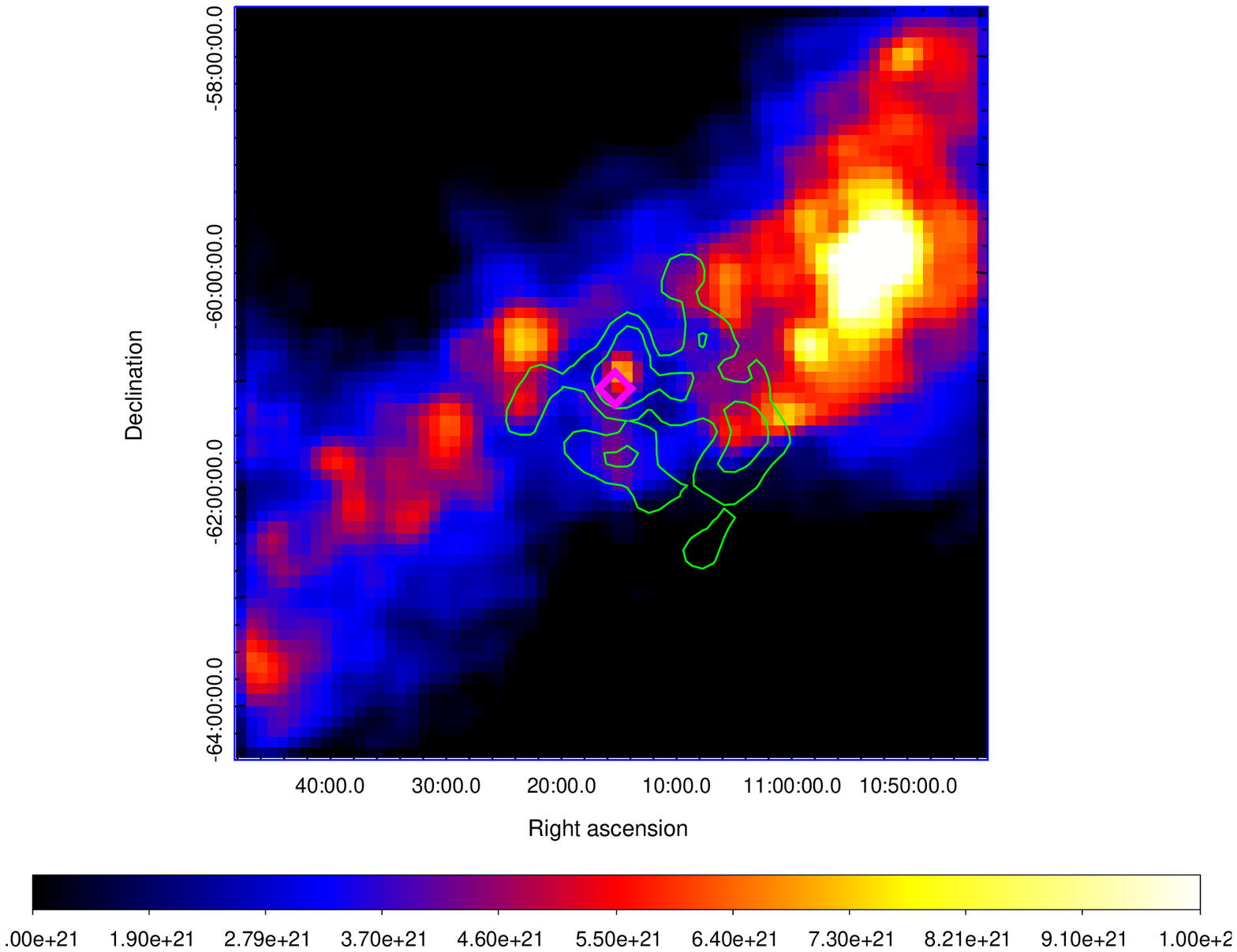}\includegraphics[width=0.3\linewidth]{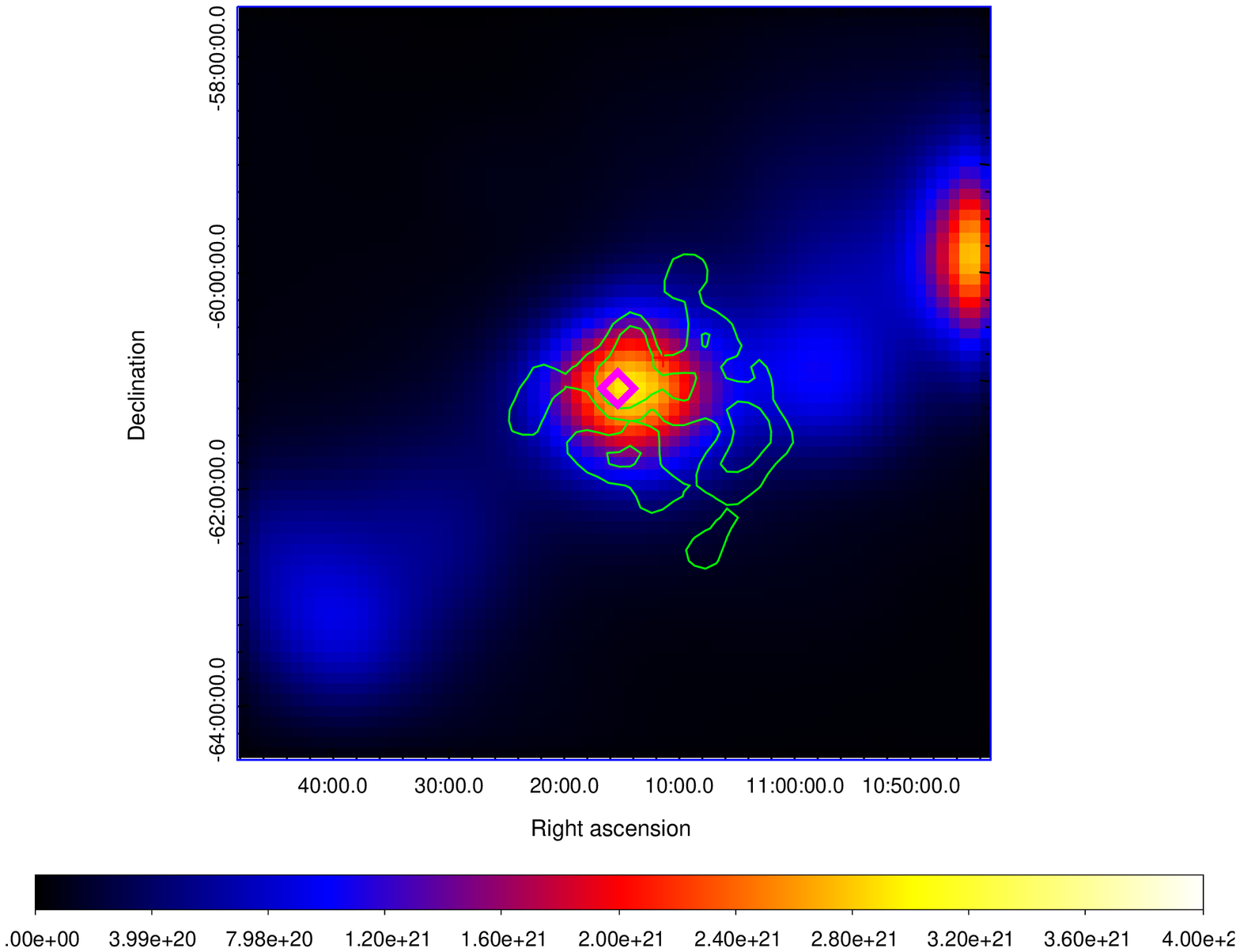}
\caption{{\it left}: Gas column derived from Planck dust opacity maps. {\it middle} :  Gas column derived from 21cm and CO data, the velocity range of $0 - 20~\rm km/s $ are chosen to integrate.   {\it right} :   The \ion{H}{ii} column derived from the Planck free-free maps.  }
\label{fig:gas}
\end{figure*}

The derived gas column of all three phases are shown in Fig.\ref{fig:gas}. We noted that the \gray emission shows good spatial correlation with the \ion{H}{ii}  map, which support the hypothesis that the emission comes from the interaction of the fresh accelerated CRs with the ambient gas in the super bubble.  The mass of gas corresponding to the diffuse emission can be determined from the gas column by assuming a distance of 7 kpc. The derived mass for different traces are listed in Table.\ref{tab:mass}. As mentioned above, the mass derived from dust opacity can be regarded as an upper limit. The lower limit of the mass can be estimated as the summation of the \ion{H}{i} and H$_2$ mass, as well as the \ion{H}{i} assuming $n_e = 10 ~\rm cm^{-3}$. Thus the total mass can be estimated in the range $2.2 \times 10^6 M_{\odot} < M <12.0 \times 10^6 M_{\odot}$. If we assume the \gray emission region is spherical in geometry and the radius can be estimated as $r = D ~\theta \sim 7.0 ~\rm kpc \times (1.1/57.29) ~\rm rad \sim130 ~\rm pc  $. Thus the average volume gas density is $10 ~\rm cm^{-3}<n_{gas}<60 ~\rm cm^{-3}$.  Taken into account the total \gray luminosity of $10^{36} ~\rm erg/s$, one derive the total CR content in this area of $2 - 10 \times 10^{49} ~\rm erg$ assuming all  \grays have a hadronic origin. This value is comparable to the total CR content in the Cygnus cocoon  estimated as  $1.3 \times 10^{49} ~\rm erg$ \citep{fermi_cygnus}.

\begin{table*}[htbp]
\caption{Gas mass derived from different tracers} \label{tab:mass} \centering
\begin{tabular}{llll}
\hline
Tracer &\vline  ~gas phase&\vline ~mass ($10^6 ~\rm M_{\odot}$)\\
\hline
Dust opacity   &\vline ~Total &\vline ~12.0 &\\

\hline
21 cm + 2.6 mm line &\vline ~\ion{H}{i}+ H$_2$ &\vline~1.7&\\
\hline
free-free intensity ($n_e = 2 ~\rm cm^{-3}$)   &\vline  ~ \ion{H}{ii} &\vline ~2.5&\\
\hline
free-free intensity ($n_e = 10 ~\rm cm^{-3}$)   &\vline  ~ \ion{H}{ii} &\vline ~0.5&\\
\hline
\end{tabular}
\end{table*}

\section{Other possible origins of the extended \gray emission}

The regions near NGC 3603 is very crowded with several pulsars and supernova remnants. The \gray emission from  pulsars typically has a cutoff at several GeV  thus  hardly can contribute to the hard \grays detected in the extended structures.  However, the pulsar wind nebulae (PWN) and supernova remnant (SNR) can be natural contributors to hard \grays. We might expect some contribution also from the enhanced diffuse emission of this region.Furthermore, due to the crowded nature and the limited understanding of the interstellar medium in this region also introduce large uncertainty of  the modeling of the Galactic diffuse \gray background. In this case the residuals may be due to the imperfect modeling of the diffuse background. Below we investigated the three possibilities in detail.

\subsection{SNRs}
There are three SNRs located inside the \gray emission region, SNR G290.1-0.8 (MSH 11-61A) \citep{slane02},  G291.0-0.1 (MSH 11-62) \citep{slane12} and G292.2-0.5 \citep{ng12}.    %
All three SNRs has a much smaller radio size compared to the size of the detected extended \gray emission. Furthermore, the age of SNR G290.1-0.8 and SNR  G291.0-0.1 are 6000 and 1900 years, respectively. In such a short time scale the CRs cannot propagate more than 100~pc even if we assume a diffusion coefficient of $10^{29} ~\rm cm/s^2$, which is the typical value in the Galactic plane,  near the SNRs. Indeed, due to the more turbulent environment near SNRs a much slower diffusion is predicted.  G292.2-0.5, with a age of more than $2\times 10^4$ years, is old enough for the accelerated CRs to occupy the \gray emission region.

\subsection{PWNs}
There are in total 8 pulsars which coincide with the \gray emission region.  5 of them are old with a rather low spin down luminosity ($<10^{34}~\rm erg$). The remaining three are PSR J1105-6107, J1112-6103 and J1119-6127.  Neither of them can produce the extended \gray emission in their wind nebulae. First of all, the size of the structure, which is more than 100 pc, can hardly be produced by young pulsars. Thus PSR J1119-6127 with a age of only 1600 years \citep{camilo00} can be excluded.  The PWN of PSR J1105-6107 has been detected in x-ray \citep{Gotthelf98} and radio \citep{Stappers99}, but with an angular size which is order of magnitude smaller.  \citet{prinz15} have performed a x-ray search for all the pulsars and found no evidence towards the direction of PSR J1112-6103.   We cannot formally rule out a PWN from a yet undiscovered pulsar, in particular in these crowded region, but the correlation between the \gray emission and the \ion{H}{ii} structure in NGC 3603 favours a CR origin. 

\subsection{imperfect modeling of diffuse background}
\begin{figure*}
\centering
\includegraphics[width=0.8\linewidth]{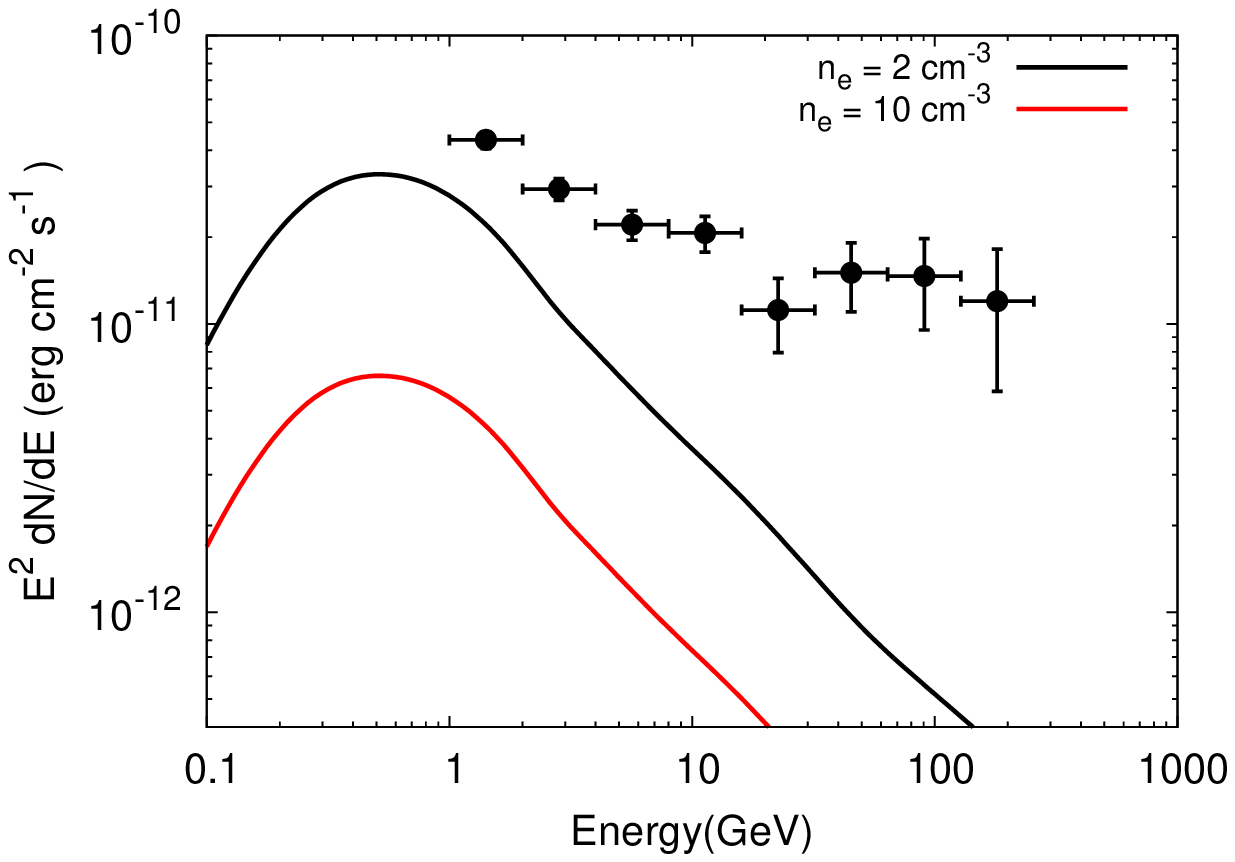}
\caption{The SEDs of the extended emission towards NGC 3603  together with the predicted \gray emission in the \ion{H}{ii} region assuming assuming the CR density therein is identical to the  density in the solar system measured by AMS-02 \citep{ams02proton}. The  \ion{H}{ii} density are derived by using equation.5 of \citet{sodroski97} and free-free intensity from \citet{planck15-10}. The electron density $n_e$ are chosen to be $2 ~\rm cm^{-3}$ and $10 ~\rm cm^{-3}$, as shown in black and red curve, respectively. 
}
\label{fig:spe_cr}
\end{figure*}

The Galactic diffuse \gray background of Fermi LAT is produced by using GALPROP code \citep{galprop}.  In the previous analysis various point sources concentrated in the galactic plane has been founded and regarded to be related with the imperfect modelling of the galactic diffuse background, which as labeled as "c" in the Fermi catalog \citep{3fgl}.  Near NGC 3603 there are no such sources. On the other hand, in the Fermi Galactic diffuse backgound models, the \grays associated with ionised gas are not included \citep{fermi_diffuse}. As mentioned in last section, NGC 3603 is the most massive \ion{H}{ii} region in the Galaxy. Thus the ignorance of   \ion{H}{ii} gas may underestimate the \gray flux towards this direction significantly. However, the hard spectrum with an index of 2.3 is not compatible with the Galactic diffuse \gray background which has an index of 2.7, which is mainly contributed by the CR interaction with gas. We plot in Figure.\ref{fig:spe_cr} the SED of NGC 3603 together with predicted \gray emission in the \ion{H}{ii} region assuming the CR density therein is identical to the  density in the solar system measured by AMS-02 \citep{ams02proton}. The observed flux is twice higher than the predicted one at 1~GeV and nearly 2 orders of magnitude higher at 200~GeV.  Thus the extended emissions are  unlikely to related  to the diffuse background components.

\section{Reanalysis of PASS 8 data on Cygnus cocoon}
\begin{figure*}
\centering
\includegraphics[width=0.8\linewidth]{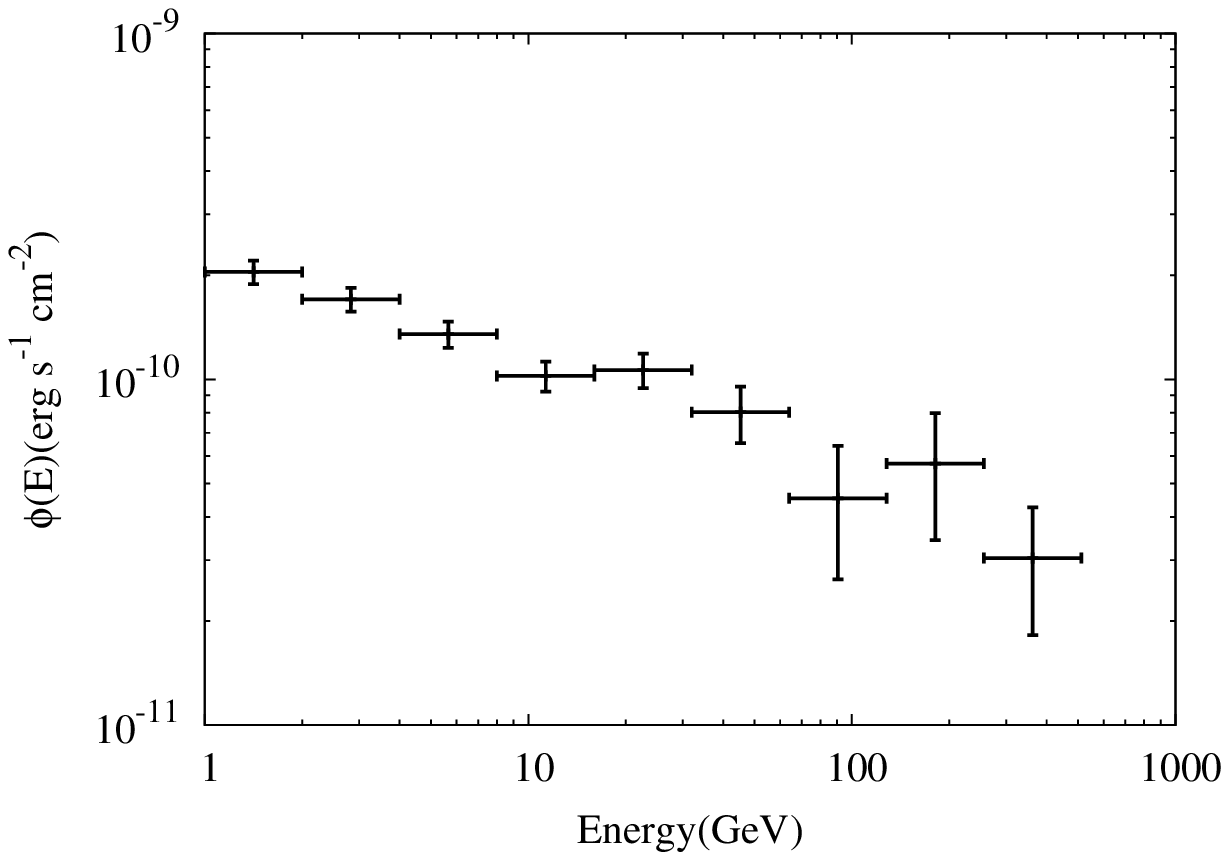}
\caption{The SEDs of cygnus cocoon.
}
\label{fig:cyg}
\end{figure*}

Cygnus cocoon \citep{fermi_cygnus} is the first detected extended \gray emission near young star clusters. The \gray emission has a hard spectrum with a photon index of -2.1 up to  100~ GeV and and has a n angular extension of   $2^{\circ}$.  The hard spectrum of Cygnus cocoon reported by \citet{fermi_cygnus} shows a similarity with NGC 3603. To take advantage of the accumulated exposure and developed understanding of instrument response, we reanalysis the region using 7 years  PASS 8 Fermi LAT data (MET 239557417 -- MET 455067824) with the latest LAT analysis software package \emph{v10r0p5}.   Cygnus cocoon is already included in the 3FGL template and we use the spatial templates therein. We divided the energy range $1000~ {\rm MeV} - 500~{\rm GeV}$ into 9 logarithmically spaced bands and applied \emph{gtlike} to each of these bands.  The derived SED of Cygnus cocoon are shown in Fig.\ref{fig:cyg}.   The spectrum above 1 GeV can be well fitted with a power law with a photon index of 2.3. The detected spectrum extends to 500 GeV, without a clear sign of cutoff.  The corresponding CR energy is up to 10~ TeV if the \grays are produced by CR interaction with gas.    
 
 \section{Discussion and conclusion}
Originally, \citet{binns05} and \citet{rauch09}, based on the composition of CRs, proposed that  Galactic CRs are produced in the supper bubbles close to OB associations/star-clusters. The recent observations of $^{60}Fe$ in CRs provide a new support of this hypothesis \citep{binns16}.  Furthermore, the measurement of \gray emissivities reveal a similar radial distribution of CRs with OB stars \citep{fermi_diffuse,yang16}.  If CRs are accelerated in such environment, high energy gamma-rays are expected from the interactions of the fresh accelerated CRs with the ambient gas. However up to now the only \gray detection towards these structures are Cygnus cocoon \citep{fermi_cygnus}. Here, we report a statistically significant detection of an extended gamma-ray signal from the direction of another  star burst region, NGC 3603.  Like the Cygnus cocoon, the spectrum of this source is hard, and extends up to 500 GeV. We argue that the most likely origin of the detected emission is the interactions of CRs accelerated by stellar winds with the ambient gas. . 

In addition to the Cygnus cocoon and NGC 3603, there are a few more similar objects like Westerlund 1, Westerlund 2, RSSG1, RSSG2, RSSG3 \citep[for a review, see][]{zwart10}, which can be consider as sites of CR accelerations and therefore as potential extended \gray sources. Interestingly, the first three of these objects have been reported as TeV gamma-r ay emitters \citep{hess_w1, hess_w2, hess_rsgc1}, thus it is likely that particles in these objects are accelerated to multi-TeV energies.  In this regard a principal question is whether these objects can operate also as PeVatrons, i.e. whether they can provide the bulk of the locally observed CRs up to the so-called knee around 1 PeV.  

The most straightforward and unambiguous answer to this question would be the detection of  gamma-rays extending with a hard energy spectrum to energies well beyond 10 TeV  Apparently,  because of the limited detection area, the Fermi LAT observations cannot offer such measurements. In this regard the Atmospheric Cherenkov Telescope Arrays with their huge detection areas and adequate  angular and energy resolutions, are 
powerful tools for the search and study of cosmic PeVatrons. 
Remarkably, multi TeV-gamma-ray emission  with a hard spectrum has been reported  from 
Westerlund 1 by by the HESS collaboration \citep{hess_w1} which has been interpreted by \citet{bykov14} as an indication of acceleration of protons in that object to PeV energies. However, this tentative results needs further confirmation. 

The forthcoming  Cherenkov Telescope Array (CTA), and to some extent  also the water Cherenkov  particle detectors like HAWK and LHAASO,   are well designed for such studies.

.  
\bibliographystyle{aa}
\bibliography{ngc3603}
\end{document}